{\newcounter{numberofpages} \setcounter{numberofpages}{17} 
 
\documentclass[11pt]{article} 
\usepackage{graphicx} 
\usepackage[tbtags]{amsmath} 
\usepackage{amsthm,amsfonts,amssymb,verbatim,cite} 
\makeatletter\let\over\@@over\makeatother 
\setcounter{page}{1} 
\newcounter{finalpage} \setcounter{finalpage}{\thepage} 
\addtocounter{finalpage}{\value{numberofpages}} 
\addtocounter{finalpage}{-1} 
\typeout{****************************************************************} 
\typeout{*      Notes pages: \thepage\space --- \arabic{finalpage}      *} 
\typeout{****************************************************************} 
\thispagestyle{empty} 
\setlength{\textwidth}{5in} 
\setlength{\textheight}{8in} 
\setlength{\topmargin}{-5mm} 
\advance\baselineskip 0pt plus 1.5pt 
\advance\parskip 0pt plus 2pt 
\makeatletter\@addtoreset{equation}{section}\makeatother 
\def\theequation{\arabic{section}.\arabic{equation}} 
\hyphenation{Thir-ring Za-mo-lod-chi-kov ho-lo-mor-phic cur-rent mar-gi-nal}
\newcommand{\TitAuthAffil}[3]{% 
\begin{center} 
\Huge #1 
\end{center} 
\begin{center} 
\vspace{0.5cm} 
{\large #2}\\ 
\vspace{0.5cm} 
{\it #3} 
\end{center} 
\vspace{1cm} 
}

\newcommand{\less}{\scriptscriptstyle <}
 
\begin{document} 
 
\rightline{CLNS 99/1622}

\TitAuthAffil{Integrable Marginal Points in the $N$-Cosine Model}% 
{Bogomil Gerganov}% 
{Newman Laboratory of Nuclear Studies \\ 
	Cornell University \\ 
	Ithaca, NY 14853 --- USA}

\begin{abstract} 
The integrability of the $N$-cosine model, a $N$-field generalization of 
the sine-Gordon model, is investigated. We establish to first order in 
conformal perturbation theory that, for arbitrary $N$, the model possesses 
a quantum conserved current of Lorentz spin 3 on a submanifold of the 
parameter space where the interaction becomes marginal. The integrability of 
the model on this submanifold is further studied using renormalization 
techniques. It is shown that for $N$= 2, 3, and 4 there exist special points 
on the marginal manifold at which the $N$-cosine model is equivalent to models
of Gross-Neveu type and therefore is integrable. In the 2-field case we 
further argue that the points mentioned above exhaust all integrable cases 
on the marginal submanifold. 
\end{abstract} 

%\rightline{05/26/1999}
 
\section{Introduction} 
The $N$-cosine model is a $N$-field generalization of the sine-Gordon 
model (SG). Multi-field generalizations of SG model have been previously 
introduced by Shankar\cite{Sh}, Bukhvostov and Lipatov\cite{BL}, 
Fateev\cite{F1,F2}, and recently by Baseilhac et al.\cite{multi:complex} and by
Saleur and Simonetti \cite{SS}. In particular, the 2-field case \cite{BL} has 
drawn closer attention in the subsequent works of Ameduri and 
Efthimiou\cite{AE}, Lesage et al. \cite{LSS1,LSS2}, Ameduri et al.\cite{AEG}, 
and Saleur\cite{long:delayed}. The boundary double-cosine model has found 
recent applications in condensed matter physics in describing tunneling 
effects in quantum wires\cite{LSS1,LSS2}.

In this work we focus on a model of $N$ boson fields, whose perturbation to the
free Lagrangian is a product of $N$ cosines, i.\ e.\ the Euclidean action of 
the model is:
\begin{equation} 
S = \frac{1}{4\pi} \int d\tau dx \left[  
\frac{1}{2} \sum_{i=1}^N {\left( \partial_\mu \Phi_i \right)}^2 
+\lambda \prod_{i=1}^N \cos({\beta}_i \Phi_i) \right] \ . 
\label{N:cos} 
\end{equation} 
 
	A calculation to first order in conformal perturbation theory 
(CPT)\cite{CPT} showed that the model (\ref{N:cos}) possesses a conserved 
current of Lorentz spin 3, when the couplings lie on the submanifold of the 
parameter space determined by the constraint
\begin{equation}
\beta_1^2 + \beta_2^2 + \ldots + \beta_N^2 = 2 \ . 
\label{T4:manifold}
\end{equation}  
The results of the above analysis are discussed in Section 2.

	The weakness of the conformal perturbation theory in the limit where 
the interaction becomes marginal\cite[p.650]{CPT} requires further 
investigation of the integrability of the model on the submanifold 
(\ref{T4:manifold}). By 
performing renormalization group analysis, conducted using the 
technique of \cite{Z:beta}, we showed that at special points on the marginal 
manifold  the model (\ref{N:cos}) can be written as a current-current 
perturbation of the level 1 Wess-Zumino-Witten (WZW) model based on some Lie 
algebra $\mathfrak{g}$. As discussed in \cite{BlC}, such models are equivalent 
to the $\mathfrak{g}$-invariant Gross-Neveu models, known to be 
integrable\cite{GN1,GN2}. Marginal models of the above type have Yangian 
symmetry and their exact $S$-matrix can be computed. A similar behavior in the 
marginal limit has been observed before by \cite{BlC} in the SG model 
(current-current perturbation of the level 1 $\mathfrak{su}{\scriptstyle (2)}$ 
WZW) and in affine Toda theories.

	In Section 3 we discuss the double-cosine model ($N$=2 case) in more
details. In addition to the previously known integrable cases in which the 
double-cosine model reduces trivially to the marginal SG, we establish the 
presence of new integrable points on the marginal manifold where the model 
can be written as a current-current perturbation of the level 1
$\mathfrak{su}{\scriptstyle (3)}$ WZW. Renormalization 
group arguments then allowed us to further argue that the aforementioned 
$\mathfrak{g}$-symmetric points exhaust all the integrable cases on the 
marginal submanifold. 
  
	Finally, in section 4, we generalize our analysis to arbitrary $N$ and
find integrable points for $N$=3 (current-current perturbation of the level 1 
$\mathfrak{su}{\scriptstyle (4)}$ WZW) and for $N$=4 (current-current 
perturbation of the level 1 $\mathfrak{so}{\scriptstyle (8)}$ WZW). Some 
calculational details are given in the Appendices.

	With this paper we hope to conclude the study of the integrability of
the models involving interactions of the multi-cosine type (\ref{N:cos}) and 
to place these QFTs in the larger context of known 1+1 dimensional 
integrable models by revealing their relationship to the imaginary coupling 
Toda, Gross-Neveu, and sine-Gordon models.

\section{Integral of Motion of Lorentz Spin 3} 
 
	In this section we investigate the presence of conserved quantities 
of higher Lorentz spin in the $N$-cosine model (\ref{N:cos}) using the 
technique 
of perturbed conformal field theory\cite{CPT}. By treating a 2D QFT as a  
perturbed CFT, it is possible to study which, if any, of the infinite number 
of conservation laws in CFT survive the perturbation. 
Zamolodchikov's paper\cite{CPT} also provides us with an easy way for 
computing the conserved current densities explicitly. 

	Including all possible local fields of dimension 4 which respect the 
symmetries of the action (\ref{N:cos}), we propose the following 
Ansatz\footnote{%
	See also \cite{F1,F2,LSS2,AEG}.}
for $T_4$ in the $N$-field case:
\begin{equation} 
T_4 =\ \sum_{i=1}^N \left[ a_i {\left( \partial^2_z \phi_i \right)}^2 + \  
 		 	 b_i {\left( \partial_z \phi_i \right)}^4 \right] 
+ \sum_{i<j} c_{ij} {\left( \partial_z \phi_i \right)}^2 
	     	    {\left( \partial_z \phi_j \right)}^2 \ .
\label{T4} 
\end{equation} 
 
	In CFT $T_4$ is a holomorphic function and $\partial_{\bar{z}}T_4 = 0$.
In the perturbed QFT that is no longer true and we can compute 
$\partial_{\bar{z}}T_4$, using Zamolodchikov's 
formula\cite[{eq.\ (3.14)}]{CPT}: 
\begin{equation} 
\partial_{\bar{z}}T_4 = \lambda \oint_z \frac{d\zeta}{2\pi i}  
\left(\prod_{i=1}^N \cos[{\beta}_i\phi_i(\zeta,\bar{z})]\right) T_4(z) \ .
\label{Zamo} 
\end{equation} 
If the RHS of (\ref{Zamo}) can be expressed as a total $\partial_z$-derivative 
of some local operator, $\partial_z \Theta_2$, the conservation law of spin 3  
survives in the perturbed QFT and has the form
\begin{equation} 
\partial_{\bar{z}} T_4 = \partial_{z} \Theta_2 \ , 
\label{Cons:Law}
\end{equation}
$T_4$ and $\Theta_2$ being the quantum conserved densities of the spin 3 
integral of motion. Our goal was to find all the conditions (if any) on the 
couplings ${\beta}_i$, $i=1,2,...,N$ for which (\ref{Cons:Law}) holds.

	As a result of the calculation one finds that if the form (\ref{T4}) 
is assumed, the spin 3 current is conserved if and only if 
\begin{equation} 
\beta^2 \equiv \sum_{i=1}^N {\beta}^2_i = \ 2 \ , ~~~ {\rm for} \ N \ge 3 \ .
\label{Manifold}
\end{equation} 

	It is interesting to compare the result for $N \ge 3$ with the 
($N$=2)-case. As shown in \cite{LSS1} and \cite{AEG}, a quantum conserved
current (to first order in CPT) exists on 3 distinct submanifolds in 
$(\beta_1,\beta_2)$-parameter space:
\begin{equation} 
	\begin{align} 
{\beta}^2_1 - {\beta}^2_2 = \ & 0 \ , 
\label{2sG} \\ 
{\beta}^2_1 + {\beta}^2_2 = \ & 1 \ , 
\label{Saleur1} \\ 
{\beta}^2_1 + {\beta}^2_2 = \ & 2 \ . 
\label{Saleur2} 
	\end{align} 
\end{equation}
The first manifold (\ref{2sG}) is trivial: when ${\beta}^2_1 = {\beta}^2_2$, 
the double cosine model decouples into 2 sine-Gordon models and, of course, 
is integrable both classically and quantum mechanically. On the second manifold
(\ref{Saleur1}) the exact $S$-matrix has been found by Lesage et 
al.\cite{LSS2}, using the method of non-local charges. The integrability of 
the model on the third manifold (\ref{Saleur2}), to the best of our knowledge, 
has not been thoroughly investigated. It is curious that only the manifold 
(\ref{Saleur2}) generalizes to the case with arbitrary $N$. Our calculation 
also showed that for $N \ge 3$, $T_4$ is not conserved on $\beta^2 = 1$ and 
not even when the squares of all the couplings are equal\footnote{%
	The latter is not as surprising as it may seem because, unlike the 
	($N$=2)-case, 
	for $N \ge 3$ the Lagrangian does not decouple trivially into a sum of 
	sine-Gordon Lagrangians even when all the couplings are equal.}.

	The fact that $T_4$ is conserved on the manifold (\ref{Manifold}) for 
arbitrary $N$ is encouraging. One could hope that it is due to some yet 
undiscovered symmetry of the theory (\ref{N:cos}). The result is, however, 
challenged by the following dimensional argument: the quantity 
$\beta^2$ (sum of the couplings squared) is exactly the dimension of the 
perturbing operator in (\ref{N:cos}) and when $\beta^2 = 2$, this operator 
becomes marginal ($[\lambda]=0$). As a result Zamolodchikov's counting 
argument\cite[p.650]{CPT} is weak in this case and first order CPT can no 
longer be claimed to give exact results. Therefore we need to use other 
methods to investigate further the integrability of the model on the marginal 
manifold.

\section{Renormalization on the Marginal Manifold and Integrable Points for 
$N=2$}

	Let us consider a conformal field theory perturbed by some marginal 
operators $\mathcal{O}_i(z,\bar{z})$:
\begin{equation} 
	S = S_{\rm CFT} + \frac{1}{4\pi} \int d^2 z \sum_i \lambda^i 
\mathcal{O}_i(z,\bar{z}) \ . 
\label{CPT:marginal} 
\end{equation} 
Zamolodchikov has shown \cite{Z:beta} that the beta functions for the 
couplings $\lambda^i$ can be computed by examining the singularities in 
the operator product expansions (OPEs) of the perturbations 
$\mathcal{O}_i(z,\bar{z})$ and by regularizing them through introducing a 
cutoff $\mu$. For an OPE of the type
\begin{equation}
	\mathcal{O}_j(w,\bar{w})\mathcal{O}_k(z,\bar{z}) \sim C^i_{jk} 
\frac{\mathcal{O}_i(z,\bar{z})}{{|w-z|}^2}
\label{OPE:pert}
\end{equation}
the divergence is logarithmic and the beta functions to second order are
\begin{equation}
	\beta_{\lambda^i} \equiv \frac{d\lambda^i}{d\log\mu} 
= \frac{1}{2} C^i_{jk} \lambda^j\lambda^k + o^3(\lambda^i) \ .
\label{beta}
\end{equation}
If the OPEs of the perturbations in (\ref{CPT:marginal}) do not close on the 
set $\left\{\mathcal{O}_i(z,\bar{z})\right\}$, the action 
(\ref{CPT:marginal}) does not define a consistent QFT. In this case 
renormalization requires that new operators be added to the Lagrangian until
$\left\{\mathcal{O}_i(z,\bar{z})\right\}$ becomes closed OPE algebra of 
the type (\ref{OPE:pert}). We refer to this procedure by saying that 
`new terms are generated under renormalization'.

	In the following discussion we shall focus on the particularly 
interesting case when the perturbing operator is of current-current type: 
\begin{equation}
	S = S_{\rm CFT} + \frac{\lambda}{4\pi} \int d^2 z 
\mathcal{O}(z,\bar{z}) \ , ~~~~~
\mathcal{O}(z,\bar{z}) = \sum_{a=1}^{\rm dim(\mathfrak{g})} 
	J_a(z)\overline{J}_a(\bar{z}) \ ,
\label{CC}
\end{equation}
where ${\left\{J_a(z)\right\}}_{a=1}^{\rm dim(\mathfrak{g})}$ are field 
representations of the generators of some Lie algebra $\mathfrak{g}$ in terms 
of bosonic vertex operators. $J_a(z)$'s satisfy the OPE\footnote{
	This OPE holds in general for a current algebra of level $k$. In this 
	paper, however, $k$ is always 1.}
\begin{equation}
J_a(w)J_b(z) \sim 
	\frac{k^2 \delta_{ab}}{{(w-z)}^2} +
	\frac{f_{ab}^c J_c(z)}{w-z} + {\rm Reg.}
\label{OPE:g}
\end{equation}
and, similarly, for $\overline{J}_a(\bar{z})$. It is then easy to show that 
\begin{equation}
	\mathcal{O}(w,\bar{w})\mathcal{O}(z,\bar{z}) \sim C_{AD} 
\frac{\mathcal{O}(z,\bar{z})}{{|w-z|}^2} \ , ~~~~~
\beta_{\lambda} \equiv \frac{d\lambda}{d\log\mu} = \frac{1}{2} C_{AD} 
\lambda^2 + o^3(\lambda)\ ,
\label{CC:beta}
\end{equation}
where $f_{ab}^c$ are the structure constants and $C_{AD}$ is the Dynkin index 
of the algebra $\mathfrak{g}$.

	Let us now apply the above renormalization technique to the model 
(\ref{N:cos}) with $N=2$. The OPE of the perturbing operator with itself 
is\footnote{In the CFT $\Phi_i(z,\bar{z}) = \phi_i(z) + 
	\overline{\phi_i}(\bar{z})$, $i=1,2$. To first order in PT,  
	$\partial_z\phi(z,\bar{z}) = \partial_z\Phi(z,\bar{z})$, 
	$\partial_{\bar{z}}\overline{\phi}(z,\bar{z}) = 
	\partial_{\bar{z}}\Phi(z,\bar{z})$ is still true even away from the 
	conformal point.}%
:
\begin{equation}
	\begin{aligned}
~ & \prod_{i=1}^2 \cos[\beta_i \Phi_i(w,\bar{w})]\prod_{i=1}^2 \cos[\beta_i 
\Phi_i(z,\bar{z})] \\
~ & ~~~ \sim \frac{1}{4} \left\{ 
\frac{1}{{|w-z|}^2} 
	\left[ \beta_1^2 :\partial_z\Phi_1(z,\bar{z})
		          \partial_{\bar{z}}\Phi_1(z,\bar{z}):
	     + \beta_2^2 :\partial_z\Phi_2(z,\bar{z})
		          \partial_{\bar{z}}\Phi_2(z,\bar{z}):
\right] \right.
\\[15pt]
~ & ~~~~~~~~~ + {|w-z|}^{2(\beta_1^2-\beta_2^2)} 
	\sum_{a,b=\pm} e^{ia\beta_1[\Phi_1(w,\bar{w})+\Phi_1(z,\bar{z})]}
		       e^{ib\beta_2[\Phi_2(w,\bar{w})-\Phi_2(z,\bar{z})]}
\\
~ & ~~~~~~~~~ \left. + {|w-z|}^{-2(\beta_1^2-\beta_2^2)} 
	\sum_{a,b=\pm} e^{ia\beta_1[\Phi_1(w,\bar{w})-\Phi_1(z,\bar{z})]}
		       e^{ib\beta_2[\Phi_2(w,\bar{w})+\Phi_2(z,\bar{z})]}
\right\} \ ,	
	\end{aligned}
\label{OPE}
\end{equation}
where the couplings $\beta_i$ are subject to the constraint 
$\beta_1^2+\beta_2^2=2$ (the marginal submanifold).

\begin{figure}[htb] 
\hspace{-7.5mm}
\includegraphics[width=8cm]{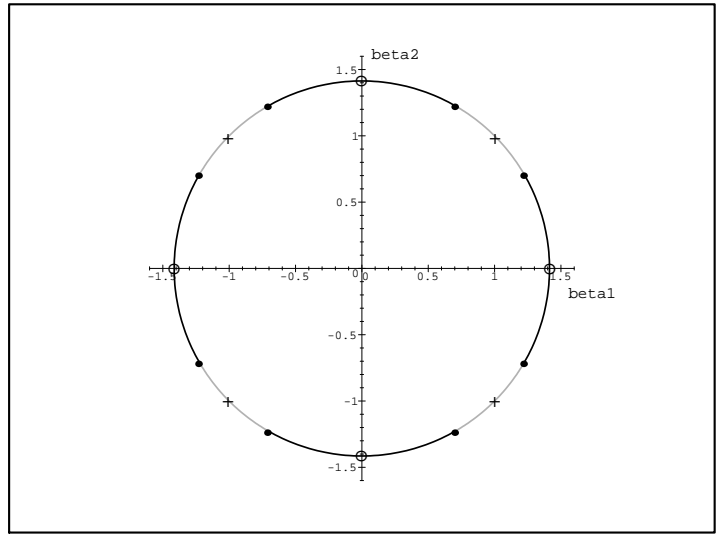} 
\vspace{-2cm}
\caption{The marginal circle.} 
\label{Figure1} 
\end{figure}

	The first term of (\ref{OPE}) leads to a logarithmic singularity like 
the one described in eqs.\ (\ref{OPE:pert}), (\ref{beta}). The type of the 
singularities in the last two terms of (\ref{OPE}) depends on the value of
the quantity $|\beta_1^2-\beta_2^2|$. There are 5 distinct cases:
\begin{equation}
	\begin{aligned}
	1. ~&~   & & |\beta_1^2-\beta_2^2|  &=&  0 \ ; \\ 
	2. ~&~ 0 &<& |\beta_1^2-\beta_2^2|  &<&  1 \ ; \\
	3. ~&~   & & |\beta_1^2-\beta_2^2|  &=&  1 \ ; \\
	4. ~&~ 1 &<& |\beta_1^2-\beta_2^2|  &<&  2 \ ; \\
	5. ~&~   & & |\beta_1^2-\beta_2^2|  &=&  2 \ .  
	\end{aligned}
\label{Cases}
\end{equation}
The regions on the marginal manifold specified by the different constraints 
in (\ref{Cases}) are pictured on Figure \ref{Figure1}. They are denoted as 
follows: the 4 points corresponding to Case 1 by ``+''; the 4 segments
corresponding to Case 2 by light grey lines; the 8 points corresponding to 
Case 3 by ``$\bullet$''; the 4 segments corresponding to Case 4 by solid 
black lines; and the 4 points corresponding to Case 5 by ``$\circ$''.

	In Case 1 and Case 2 the last two terms of (\ref{OPE}) are regular. 
In Case 1 the couplings $\beta_1$ and $\beta_2$ are equal and the model 
(\ref{N:cos}) decouples into two sine-Gordon models\footnote{See also eq.\ 
(\ref{2sG}) and the discussion that follows.}. In Case 2 the double-cosine 
model is non-trivial but no new terms are brought in by renormalization. In 
Case 3 one of the last two terms of (\ref{OPE}) is logarithmically divergent 
and 
contributes to the beta function which is of the type (\ref{beta}). This case 
is the most interesting one and we will discuss it in some more details. In 
Case 4 the last two terms give rise to power-law divergences which lead to a 
generation of a mass term under renormalization. In Case 5 one of the 
couplings vanishes while the other has a value of $\pm\sqrt{2}$. In this case 
(\ref{N:cos}) is reduced to a single SG model at its marginal point and a free
boson field. (The marginal limit of the SG model is discussed in details in 
\cite{BlC}.)

	\subsection{The Model as a Current-Current Perturbation}

	In this subsection we will show that the double-cosine interaction can 
be written as a current-current perturbation at the marginal points where the 
quantity $|\beta_1^2-\beta_2^2|$ has integer values --- these are cases 1, 3, 
and 5 of (\ref{Cases}). In each of these cases the values of $\beta_1$ and 
$\beta_2$ are such that the double-cosine perturbation can be written in the 
form  
\begin{equation}
	\mathcal{O}(z,\bar{z}) 
= \prod_{i=1}^2 \cos[\beta_i \Phi_i(z,\bar{z})] 
= \sum_{\alpha_j \ : \ {\rm some}} 
  E^{\alpha_j}(z)\overline{E}^{-\alpha_j}(\bar{z}) \ ,
\label{2cos:CC}
\end{equation}
where $\left\{ \alpha_j \right\}$ are some (but not all) of the roots and 
$E^{\alpha_j}(z)$ are the corresponding generators of some Lie algebra 
$\mathfrak{g}$ in the Cartan-Weyl basis\footnote{%
	In order to write the double-cosine perturbation in terms of the 
	generators $E^{\alpha_j}(z)$ we need to use the field representation 
	of $\mathfrak{g}$ in terms of bosonic vertex operators:
	\begin{equation}
	E^{\alpha_j}(z) = e^{i\alpha_j \cdot \vec{\phi}}(z) \ , ~~~~~
	\overline{E}^{\alpha_j}(\bar{z}) = e^{i\alpha_j \cdot 
		\vec{\overline{\phi}}}(\bar{z}) \ ,
	\label{Vert:Rep}
	\end{equation}
	where $\vec{\Phi} = (\Phi_1, \ldots, \Phi_N)$ and $\alpha_j$ is a root 
	of $\mathfrak{g}$ (by tradition, the vector sign over $\alpha_j$ is 
	omitted). For a detailed discussion of the bosonic vertex 
	representation of Lie algebras and the OPEs of the vertex operators 
	see for instance \cite{CFT,ALA:QG}.}.
As we will show below, the generators corresponding to all roots as well as 
the Cartan generators of $\mathfrak{g}$ are generated under renormalization, 
thus leading to a current-current perturbation to the level 1 $\mathfrak{g}$
current algebra\footnote{%
	Also called the $\mathfrak{g}$ Wess-Zumino-Witten (WZW) model at level 
	1.}:
\begin{multline}
S = S_{\rm CFT} \\
      +	\frac{\lambda}{4\pi} \int \hspace{-1mm} d^2 z \hspace{-1mm} \left\{
	\sum_{\alpha_j>0} \hspace{-1mm} \left[
		E^{\alpha_j}(z)\overline{E}^{-\alpha_j}(\bar{z}) +
		E^{-\alpha_j}(z)\overline{E}^{\alpha_j}(\bar{z}) \right]
      +	\hspace{-1mm} 
	\sum_{i=1}^{\rm rank(\mathfrak{g})} H^i(z)\overline{H}^i(\bar{z}) 
\hspace{-1mm} \right\} \hspace{-0.65mm} .
\label{g:CC}
\end{multline}

	Let us first consider Case 1 of eq.\ (\ref{Cases}). The condition 
$|\beta_1^2-\beta_2^2|=0$ together with the constraint that $\beta_i$ lie on 
the marginal manifold, $\beta_1^2+\beta_2^2=2$, specifies 4 points in the
parameter space:
\begin{equation}
	\beta_1^2=\beta_2^2=1 \ .
\label{Points:SU2xSU2}
\end{equation}
At these points the model decouples into two SG models\footnote{The 
	$\mathfrak{su}{\scriptstyle (2)} \oplus 
	\mathfrak{su}{\scriptstyle (2)}$-invariance of the double-cosine model
	at these points has also been discussed by Shankar in 
	\cite{Sh}.}, 
each at its marginal point. As shown for instance in \cite{BlC}, the SG model 
in the marginal limit is equivalent under renormalization to a current-current 
perturbation of the $\mathfrak{su}{\scriptstyle (2)}$ current algebra and 
also to the $\mathfrak{su}{\scriptstyle (2)}$ Gross-Neveu model.

	Next, let us consider Case 3 of eq.\ (\ref{Cases}). The constraints 
$|\beta_1^2-\beta_2^2|=1$ and $\beta_1^2+\beta_2^2=2$ specify 8 points in 
the parameter space:
\begin{equation}
	\beta_1^2=\frac{3}{2} \ , ~~ \beta_2^2=\frac{1}{2} \ ; ~~~~~~~
	\beta_1^2=\frac{1}{2} \ , ~~ \beta_2^2=\frac{3}{2} \ .
\label{Points:SU3}
\end{equation}
The RHS of the generic OPE (\ref{OPE}) then reads:  
\begin{equation}
\frac{1}{4} \frac{1}{{|w-z|}^2} \left\{ 
	\sum_{i=1}^2 \beta_i^2 \partial_z\Phi_i(z,\bar{z})
		          \partial_{\bar{z}}\Phi_i(z,\bar{z})
	+ 2\cos[2\beta_{\less} \Phi_{\less}(z,\bar{z})] \right\} \ .
\label{OPE:SU3}
\end{equation}
The full renormalized action at the marginal points (\ref{Points:SU3}) is 
\begin{multline} 
S = \frac{1}{4\pi} \int d\tau dx \left\{  
\frac{1}{2} \sum_{i=1}^2 {\left( \partial_\mu \Phi_i \right)}^2 \right. \\
+ \left. \lambda \left[
      	\sum_{i=1}^2 \beta_i^2 \partial_z\Phi_i \partial_{\bar{z}}\Phi_i
      +	\cos({\beta}_1 \Phi_1)\cos({\beta}_2 \Phi_2)
      + \cos(2\beta_{\less} \Phi_{\less}) 
\right] \right\} \ ,
\label{S:SU3} 
\end{multline} 
where the values of $\beta_1$ and $\beta_2$ are solutions of 
(\ref{Points:SU3}), $\beta_{\less} \equiv \min(\beta_1,\beta_2)$, and 
$\Phi_{\less}$ is the field corresponding to $\beta_{\less}$. Consulting the 
vertex representations of the Kac-Moody currents of 
$\mathfrak{su}{\scriptstyle (3)}$ and their OPEs\footnote{See for instance 
	\cite{CFT,ALA:QG}.}, 
we can rewrite the $\lambda$-term of (\ref{S:SU3}) as a current-current 
perturbation\footnote{See Appendix A for details.} 
of the current algebra $\mathfrak{su}{\scriptstyle (3)}$ (the 
$\mathfrak{su}{\scriptstyle (3)}$ WZW model).
\begin{equation} 
	\frac{\lambda}{4\pi} \int d^2 z \left\{
	\sum_{\alpha_i>0} \left[
		E^{\alpha_i}(z)\overline{E}^{-\alpha_i}(\bar{z}) +
		E^{-\alpha_i}(z)\overline{E}^{\alpha_i}(\bar{z}) \right]
	+ \sum_{i=1}^{2}
		H^i(z)\overline{H}^i(\bar{z}) \right\} \ .
\label{CC:SU3}
\end{equation}
This is also equivalent\cite{BlC} to the $\mathfrak{su}{\scriptstyle (3)}$ 
Gross-Neveu model. Therefore, at the points (\ref{Points:SU3}) the 
double-cosine model is integrable and yet nontrivial since it is not 
reduced to the SG model.

	Finally, in Case 5 of eq.\ (\ref{Cases}), the constraints 
$|\beta_1^2-\beta_2^2|=2$ and $\beta_1^2+\beta_2^2=2$ specify the 4 points  
\begin{equation}
	\beta_1^2=2 \ , ~~ \beta_2^2=0 \ ; ~~~~~~~
	\beta_1^2=0 \ , ~~ \beta_2^2=2
\label{Points:SU2}
\end{equation}
where the model decouples into a free field and a single marginal SG model
(equivalent under renormalization to a $\mathfrak{su}{\scriptstyle (2)}$ 
Gross-Neveu model \cite{BlC}).

	\subsection{The Model away from the Gross-Neveu Points}

	So far we have discussed the behavior of the marginal double-cosine 
model only at the special points where the model can be written as a 
current-current perturbation. Let us now complete the renormalization group 
analysis on the entire marginal manifold. The regions that remain to be 
considered are described by Cases 2 and 4 of eq.\ (\ref{Cases}). (See Figure
\ref{Figure1}.)

	Case 2 corresponds to the regions on the marginal circle, 
$\beta_1^2 + \beta_2^2 = 2$, where the couplings also satisfy the constraint 
$0 < |\beta_1^2-\beta_2^2| < 1$ (the grey segments on Figure \ref{Figure1}). 
On these segments the only contribution to the beta function of $\lambda$ 
comes from the first term of (\ref{OPE}) while the last two terms do not give 
rise to singularities. The full renormalized action in this regime reads:
\begin{equation} 
S = \frac{1}{4\pi} \int d\tau dx \left\{  
\frac{1}{2} \sum_{i=1}^2 {\left( \partial_\mu \Phi_i \right)}^2 
+\lambda\left[ 
 \prod_{i=1}^2 \cos({\beta}_i \Phi_i) +
 \sum_{i=1}^2  \beta_i^2 \partial_z\Phi_i\partial_{\bar{z}}\Phi_i
	\right]
\right\} \ . 
\label{Reg1} 
\end{equation} 
The beta function for the coupling $\lambda$ is:
\begin{equation}
\beta_\lambda \equiv \mu\frac{d\lambda}{d\mu} = 
\frac{1}{2} \sqrt{\frac{\beta_1^4 + \beta_2^4}{2}} \lambda^2 
	+ o^3(\lambda) \ .
\label{beta:Reg1}
\end{equation}
Although the model in this regime is non-trivial and its properties could
be further investigated, the following simple argument makes integrability 
look very unlikely: In the limit $|\beta_1^2 - \beta_2^2| \longrightarrow 0$
the model is equivalent to $\mathfrak{su}{\scriptstyle (2)} \oplus 
\mathfrak{su}{\scriptstyle (2)}$ current-current perturbation (two 
decoupled SG models) and its particle spectrum consists of 4 particles 1 
soliton and 1 anti-soliton for each SG model (the fundamental representation 
of each $\mathfrak{su}{\scriptstyle (2)}$ is 
$\mathbf{2}$). In the other limit, 
$|\beta_1^2 - \beta_2^2| \longrightarrow 1$, the model is equivalent to a 
$\mathfrak{su}{\scriptstyle (3)}$ current-current perturbation and its spectrum
consists of 6 particles --- belonging to the two fundamental representations, 
$\mathbf{3}$ and $\mathbf{\bar{3}}$, of $\mathfrak{su}{\scriptstyle (3)}$. 
Therefore it is not conceivable, just by counting the degrees of freedom, 
that the particle spectrum could be smoothly deformed from the first point to 
the second.

	Case 4 corresponds to the segments on the marginal circle, 
where the couplings also satisfy $1 < |\beta_1^2-\beta_2^2| < 2$ (the black
segments on Figure \ref{Figure1}). On these 
segments, in addition to the contribution to $\beta_\lambda$ 
coming from the logarithmic singularity due to the first term of (\ref{OPE}),
it is necessary to add new counterterms to the action to cancel the 
singularity coming from one of the last two terms of (\ref{OPE}). This is a 
power-law singularity and hence, by dimensional analysis, the
coupling of this new counterterm has to be massive\footnote{In the limit 
	$|\beta_1^2-\beta_2^2| \longrightarrow 1$, the coupling $\Lambda$ also
	becomes dimensionless and the divergence in the OPE logarithmic. In 
	this limit (\ref{Reg2}) reduces to the single-coupling current-current 
	perturbation (\ref{S:SU3}).}%
. The full renormalized action reads:
\begin{multline} 
S = \frac{1}{4\pi} \int d\tau dx \left\{  
\frac{1}{2} \sum_{i=1}^2 {\left( \partial_\mu \Phi_i \right)}^2 \right. \\
\left.
+\lambda\left[ 
 \prod_{i=1}^2 \cos({\beta}_i \Phi_i) +
 \sum_{i=1}^2  \beta_i^2 \partial_z\Phi_i\partial_{\bar{z}}\Phi_i
	\right]
+\Lambda \cos(2\beta_{\less}\Phi_{\less})
\right\} \ ,
\label{Reg2} 
\end{multline} 
where $\beta_{\less} \equiv \min(\beta_1,\beta_2)$ and $\Phi_{\less}$ is the 
corresponding field. The renormalization group equations in this case are
\begin{equation}
	\begin{aligned}
\beta_\lambda \equiv \mu\frac{d\lambda}{d\mu} & = 
\frac{1}{2} \sqrt{\frac{\beta_1^4 + \beta_2^4}{2}} \lambda^2 
	+ o^3(\lambda) \\
\beta_\Lambda \equiv \mu\frac{d\Lambda}{d\mu} & = 
2\beta_{\less}^2\Lambda 
+\frac{\lambda^2}{8} \mu^{2\left(\left|\beta_1^2-\beta_2^2\right|-1\right)} 
+\frac{2\beta_{\less}^4}{\sqrt{2\left(\beta_1^4+\beta_2^4\right)}}
	\lambda\Lambda
+ o^3(\lambda,\Lambda) \ .
	\end{aligned}
\label{beta:Reg2}
\end{equation}
It can be easily checked that the massive $\Lambda$-perturbation in 
(\ref{Reg2}), generated under renormalization, breaks the conservation of 
the spin 3 current (\ref{Cons:Law}) even to first order in CPT. Therefore, 
in this region the model is not integrable.

\section{Generalization to Arbitrary $N$}

	In the previous section we completed our analysis of the 
integrability of the double-cosine model ($N$=2) on the marginal manifold by 
considering all distinct cases specified by (\ref{Cases}). Let us now try to 
generalize our analysis to the $N$-field case. We will be looking in 
particular for special points on the marginal $N$-sphere (\ref{Manifold}) 
where the $N$-cosine term can be written as a current-current perturbation to 
some current algebra $\mathfrak{g}$. 

	Expanding the cosines in terms of exponentials, we get:
\begin{equation}
\prod_{i=1}^N \cos(\beta_i \Phi_i) =
\sum_{\nu=0}^{2^N-1} e^{i\vec{\beta}_\nu \cdot \vec{\Phi}} \ ,
\label{N:vertex}
\end{equation}
where $\vec{\Phi} \equiv \left( \Phi_1, \ldots, \Phi_N \right)$ and
$\vec{\beta}_\nu \equiv \left( s_\nu^1 \beta_1, \ldots, s_\nu^N 
\beta_N \right)$ with $s_\nu^i \in \{-1,+1\}$. All combinations of 
signs, $s_\nu^i$, are present in the RHS of (\ref{N:vertex}), so that there 
are $2^N$ different vectors $\vec{\beta}_\nu$ in the $N$-field case\footnote{%
	The vectors $\beta_\nu$ can be conveniently labeled as 
	$\vec{\beta}_\nu = \left({(-1)}^{\nu^{N-1}}\beta_1, \dots, 
	{(-1)}^{\nu^0}\beta_N\right)$, where $\nu^i$ is the $i^{\rm th}$ 
	binary digit of $\nu$. For instance, for $N=3$, 
	$\beta_6 = \beta_{110} = ({(-1)}^1\beta_1, {(-1)}^1\beta_2, 
	{(-1)}^0\beta_3) = (-\beta_1, -\beta_2, \beta_3)$.}%
. Our goal is to look for special values of $\beta_i$ for which the 
vectors $\beta_\nu$ become roots of some Lie algebra $\mathfrak{g}$. 

	There are a few constraints on $\mathfrak{g}$ that can be immediately
seen:
	\begin{enumerate}
	\item $\mathfrak{g}$ must be simply-laced, since 
$(\vec{\beta}_\nu \cdot \vec{\beta}_\nu)= 2$, $\forall \nu$ and 
thus all the roots of $\mathfrak{g}$ must have equal length.
	\item $\mathbf{rank}[\mathfrak{g}]$ must be $N$, since we will be 
looking for $N$-field vertex representations of the generators of 
$\mathfrak{g}$.
	\end{enumerate}
The simply-laced Lie algebras of rank $N$ are $\mathbf{A}_N$, 
$\mathbf{D}_{(N\ge4)}$, and $\mathbf{E}_{(N=6,7,8)}$. Before trying to write
(\ref{N:vertex}) as a current-current perturbation to one of these algebras, 
let us first make the following simple considerations:

	First, since ${| \vec{\beta}_\nu |}^2$=2, each of its 
components must satisfy:
\begin{equation}
{(s_\nu^i \beta_i)}^2 = \beta_i^2 \in (0,2) \ .
\label{Interval}
\end{equation}
The ends of the above interval, 0 and 2,  are excluded because in both cases 
some of the couplings $\beta_i$ will vanish and the $N$-field model will 
trivially reduce to a lower-$N$ case. Furthermore, if $\{\vec{\beta}_\nu\}$ 
are roots of a simply-laced Lie algebra with squared norm 2, it is easily
seen from the Cartan matrix that
\begin{equation}
	\begin{aligned}
(\vec{\beta}_\nu \cdot \vec{\beta}_\lambda ) & = \pm 2 \ , 
	&~\mathrm{for} \ \nu=\pm\lambda \ ; \\
(\vec{\beta}_\nu \cdot \vec{\beta}_\lambda ) & \in \{-1,0,+1\} \ , 
	&~\mathrm{for} \ \nu\neq\pm\lambda \ .
	\end{aligned}
\label{Scal:Prod}
\end{equation}
Let us assume, for instance, that $\vec{\beta}_\nu$ and $\vec{\beta}_\lambda$
($\nu \neq \lambda$) differ only by their $p^\mathrm{th}$ components:
$s_\nu^p = - s_\lambda^p$. According to (\ref{Scal:Prod}) we have:
\begin{equation}
	(\vec{\beta}_\nu \cdot \vec{\beta}_\lambda ) 
= \sum_{i=1}^{N} \beta_i^2 - 2\beta_p^2 = n_p \in \{-1,0,+1\} \ .
\label{One:diff}
\end{equation}
Since for any $p \in \{1, \ldots, N\}$, there is a pair of vectors for which 
(\ref{One:diff}) is true, we conclude, using also (\ref{Manifold}), that 
\begin{equation}
\beta_i^2 \in \left\{ \frac{1}{2}, 1, \frac{3}{2} \right\} \ , 
~ i=1, 2, \ldots, N \ , ~~ N \ge 2 \ .
\label{Values}
\end{equation}
In other words, we can only hope to be able to write the $N$-cosine model as 
a current-current perturbation at the points on the marginal manifold whose 
coordinates satisfy (\ref{Values}). As $N$ increases, (\ref{Manifold}) 
additionally constrains the set (\ref{Values}), so we finally get:
\begin{equation}
	\begin{align}
	\mathrm{For}~ N=2: & ~~~~~
\beta_i^2 \in \left\{ \frac{1}{2}, 1, \frac{3}{2} \right\} \ , 
&~ i=1, 2 \ ; \label{N2} \\
	\mathrm{For}~ N=3: & ~~~~~
\beta_i^2 \in \left\{ \frac{1}{2}, 1 \right\} \ , 
&~ i=1, 2, 3 \ ; \label{N3} \\
	\mathrm{For}~ N=4: & ~~~~~
\beta_i^2 \in \left\{ \frac{1}{2} \right\} \ , 
&~ i=1, 2, 3, 4 \ ; \label{N4} \\
	\mathrm{For}~ N>4: & ~~~~~
\beta_i^2 \in \emptyset \ , 
&~ i=1, 2, \ldots, N \label {N+}\ .
	\end{align}
\end{equation}
Therefore, for $N>4$ the marginal $N$-cosine model cannot be written as
a current-current perturbation.

	Let us now consider in some more detail the remaining two cases,
$N$=3 and $N$=4. We shall discuss below that in both cases there exist 
specific sets of values for the couplings $\beta_i$ specifying isolated 
points on the marginal manifold (\ref{Manifold}) where the model 
(\ref{N:cos}) is equivalent under renormalization to a a current-current 
perturbation of some current algebra $\mathfrak{g}$:
\begin{equation}
      \frac{\lambda}{4\pi} \int d^2 z \left\{
	\sum_{\alpha_j>0} \left[
		E^{\alpha_j}(z)\overline{E}^{-\alpha_j}(\bar{z}) +
		E^{-\alpha_j}(z)\overline{E}^{\alpha_j}(\bar{z}) \right]
      +	\hspace{-1mm} 
\sum_{i=1}^{\rm rank(\mathfrak{g})} H^i(z)\overline{H}^i(\bar{z}) \right\} \ .
\label{CC:G}
\end{equation}
This is a model of the Gross-Neveu type with $\mathfrak{g}$-symmetry and 
such QFTs are known to be integrable. The $S$-matrices for the 
$\mathfrak{g}$-invariant models of Gross-Neveu type are known explicitly
for the classical Lie algebras \cite{GN1,GN2} and their structure is expected to 
be the same for all Lie algebras.

	In the $N=3$ case the only possible candidate for $\mathfrak{g}$ is 
$\mathbf{A}_3 = \mathfrak{su}{\scriptstyle (4)}$, since this 
is the only simply-laced Lie algebra of rank 3. There are 24 points on the 
marginal manifold (\ref{Manifold}) for which the values of the couplings 
$\beta_i$ are consistent with (\ref{N3}). These points are specified by the 
equations:
\begin{equation}
(\beta_1^2,\beta_2^2,\beta_3^2) = 
	\left( \frac{1}{2},1,\frac{1}{2} \right) \ , ~ 
	\left( 1,\frac{1}{2},\frac{1}{2} \right) \ , ~
	\left( \frac{1}{2},\frac{1}{2},1 \right) \ .
\label{Points3}
\end{equation}
At the above points, the 3-cosine action is equivalent\footnote{%
	Some details of the calculation are provided in Appendix B.}
to a current-current perturbation of the type (\ref{CC:G}) with $\mathfrak{g} =
\mathfrak{su}{\scriptstyle (4)}$. In this case $\alpha_j$ runs over the 3 
simple roots and the 3 non-simple positive 
roots of $\mathfrak{su}{\scriptstyle (4)}$ and $-\alpha_j$ are the 
corresponding negative roots. The perturbation of (\ref{CC:G}) thus contains 
the generators for all 12 roots of $\mathfrak{su}{\scriptstyle (4)}$ and the 
3 Cartan generators, which is consistent with 
{\bf rank}$[\mathfrak{su}{\scriptstyle (4)}]$=3 and 
{\bf dim}$[\mathfrak{su}{\scriptstyle (4)}]$=15.

	For $N=4$ the only possible candidates for $\mathfrak{g}$ are 
$\mathbf{A}_4 = \mathfrak{su}{\scriptstyle (5)}$ and 
$\mathbf{D}_4 = \mathfrak{so}{\scriptstyle (8)}$, the only simply-laced 
Lie algebras of rank 4. There are 16 points on the marginal 4-sphere 
(\ref{Manifold}) for which the values of the couplings $\beta_i$ are 
consistent with (\ref{N4}). Their coordinates are fixed by the equation:
\begin{equation}
(\beta_1^2,\beta_2^2,\beta_3^2,\beta_4^2) = 
	\left( \frac{1}{2},\frac{1}{2},\frac{1}{2},\frac{1}{2} \right) \ .
\label{Points4}
\end{equation}
The vectors $\vec{\beta}_\nu$ for the above values of the couplings do not 
satisfy the Cartan matrix of $\mathfrak{su}{\scriptstyle (5)}$ so we are left 
with the second candidate, $\mathfrak{so}{\scriptstyle (8)}$. 
We indeed showed\footnote{%
	Some calculational details are provided in Appendix C.} 
that at the above points the 4-cosine action is equivalent 
under renormalization to a current-current perturbation of the 
$\mathfrak{g}=\mathfrak{so}{\scriptstyle (8)}$ current algebra. $\alpha_j$ 
runs over the 4 simple roots and the 8 non-simple positive 
roots of $\mathfrak{so}{\scriptstyle (8)}$ and $-\alpha_j$ are the 
corresponding negative roots. The perturbation of (\ref{CC:G}) thus contains 
the generators for all 24 roots of $\mathfrak{so}{\scriptstyle (8)}$ and the 
4 Cartan generators, which is consistent with 
{\bf rank}$[\mathfrak{so}{\scriptstyle (8)}]$=4 and 
{\bf dim}$[\mathfrak{so}{\scriptstyle (8)}]$=28.

\section{Conclusions} 
We have completed the analysis of the integrability of the double-cosine 
model on the marginal manifold and have found that the model is integrable 
only at the points where the interaction can be written as a current-current 
perturbation to some current algebra $\mathfrak{g}$. In addition to the 
previously known points with symmetry $\mathfrak{su}{\scriptstyle (2)}$ 
(single SG) and 
$\mathfrak{su}{\scriptstyle (2)} \oplus \mathfrak{su}{\scriptstyle (2)}$ 
(2 decoupled SGs), we have also found 
$\mathfrak{su}{\scriptstyle (3)}$-symmetric points where the model is of 
Gross-Neveu type and is thus integrable. Similarly, in the $N$=3 case we
observed $\mathfrak{su}{\scriptstyle (4)}$-invariant integrable points and 
in the $N$=4 case, $\mathfrak{so}{\scriptstyle (8)}$-invariant integrable 
points. 

At last, we would like to compare the marginal limiting behaviors of the 
multi-cosine models and the imaginary coupling affine Toda theories 
\cite{BlC}. Toda theories depend on a single dimensionless coupling parameter.
When this parameter approaches the value for which the perturbing operator 
becomes marginal, the theory is equivalent to a current-current perturbation 
of the WZW model based on some Lie algebra $\mathfrak{g}$. The full current 
algebra in the Toda case is generated under renormalization by the vertex 
operators in the original action corresponding to the simple roots and the 
affine root of $\mathfrak{g}$. The multi-cosine models, on the other hand,  
depend on many coupling parameters and become marginal on an entire 
hypersphere in the parameter space. At special points 
on this hypersphere they are also equivalent to a current-current perturbation 
of the $\mathfrak{g}$-symmetric WZW model, the full current algebra being 
generated by vertex operators in the original action corresponding to the 
simple and the negative-simple roots of $\mathfrak{g}$. Therefore, even though 
in general the $N$-cosine models are very different from the affine Toda 
theories, at special marginal points they have the same limiting behavior.

\section*{Acknowledgments} 

	I would like to thank Andr\'{e} LeClair for support and advice and 
for helping me with numerous useful discussions and insights throughout 
the completion of this work. I am also thankful to Marco Ameduri and 
Zorawar Bassi for many discussions and useful ideas. 
 
\section*{Appendices: $N$-Cosine Models as Current-Current Perturbations}

\subsection*{A $~~~$ $\mathfrak{su}{\scriptstyle (3)}$ Current-Current 
Perturbation for $N$=2}

%%%%%%%%%% NUMBERING EQUATIONS WITHIN APPENDIX A %%%%%%%%% 
\makeatletter\@addtoreset{equation}{section}\makeatother 
\setcounter{equation}{0}
\def\theequation{A.\arabic{equation}} 
%%%%%%%%%%%%%%%%%%%%%%%%%%%%%%%%%%%%%%%%%%%%%%%%%%%%%%%%% 

	In this Appendix we provide some details of our calculation, showing 
that the 2-cosine model can be written as a $\mathfrak{su}{\scriptstyle (3)}$
current-current perturbation at the 8 points on the marginal 2-sphere 
(\ref{Manifold}) given by (\ref{Points:SU3}).

	The Cartan matrix and the Dynkin diagram of 
$\mathfrak{su}{\scriptstyle (3)}$ are:
\begin{equation}
\hspace{-25mm}
	\mathbf{A} =
\left(
\begin{array}{rr}
 2 & -1 \\
-1 &  2 
\end{array}	
\right) \ ,
\hspace{30mm}
\begin{picture}(0,0)(0,0)
	\setlength{\unitlength}{1mm}
	\put(0,0){\circle{2}}
	\put(1,0){\line(1,0){8}}	
	\put(10,0){\circle{2}}
	\put(-3,-5){${\scriptstyle (1,1)}$}
	\put( 7,-5){${\scriptstyle (2,1)}$}
\end{picture}
\label{Cartan:SU3}
\end{equation}
Let us take for example the point $\beta_1$=$\frac{1}{\sqrt{2}}$, 
$\beta_2$=$\sqrt{\frac{3}{2}}$. The double-cosine perturbation is expanded 
in terms of vertex operators as in (\ref{N:vertex}), where the vectors 
$\vec{\beta}_\nu$ reproduce the simple roots\footnote{%
	The roots in this and the following Appendices are not written in the 
	basis conventionally used in the literature. Since we have
	to fulfill the requirement that $\beta_i \neq 0$, we must find such 
	basis of the root space in which all components of all roots are 
	non-vanishing. All the roots here are written in such basis.} 
of $\mathfrak{su}{\scriptstyle (3)}$ and their opposite negative roots:
\begin{equation}
	\begin{aligned}
\vec{\beta}_0 & = \vec{\beta}_{00} = 
\left(  \frac{1}{\sqrt{2}},  \sqrt{\frac{3}{2}} \right) =  \alpha_1 \ ,
&~~~& \vec{\beta}_3=\vec{\beta}_{11} 		        = -\alpha_1 \ ; \\
\vec{\beta}_1 & = \vec{\beta}_{01} = 
\left(  \frac{1}{\sqrt{2}}, -\sqrt{\frac{3}{2}} \right) =  \alpha_2 \ ,
&~~~& \vec{\beta}_2=\vec{\beta}_{10} 		        = -\alpha_2 \ .
	\end{aligned}
\label{Roots:SU3}
\end{equation}
We see that the exponential form (\ref{N:vertex}) of the 2-cosine perturbation 
contains all $\mathfrak{su}{\scriptstyle (3)}$ generators corresponding to
the simple roots $\alpha_1$, $\alpha_2$ and their opposite negative roots 
$-\alpha_1$, $-\alpha_2$. The generators corresponding to the remaining
roots of $\mathfrak{su}{\scriptstyle (3)}$, $\alpha_1$+$\alpha_2$ and 
$-\alpha_1{\rm -}\alpha_2$, as well as the 2 Cartan generators of 
$\mathfrak{su}{\scriptstyle (3)}$, are generated under renormalization.

\subsection*{B $~~~$ $\mathfrak{su}{\scriptstyle (4)}$
Current-Current Perturbation for $N$=3}

%%%%%%%%%% NUMBERING EQUATIONS WITHIN APPENDIX B %%%%%%%%% 
\makeatletter\@addtoreset{equation}{section}\makeatother 
\setcounter{equation}{0}
\def\theequation{B.\arabic{equation}} 
%%%%%%%%%%%%%%%%%%%%%%%%%%%%%%%%%%%%%%%%%%%%%%%%%%%%%%%%% 

	In this Appendix we provide some details of our calculation, showing 
that the 3-cosine model can be written as a $\mathfrak{su}{\scriptstyle (4)}$
current-current perturbation at the 24 points on the marginal 3-sphere 
(\ref{Manifold}) given by (\ref{Points3}).

	The Cartan matrix and the Dynkin diagram of 
$\mathfrak{su}{\scriptstyle (4)}$ are:
\begin{equation}
\hspace{-25mm}
	\mathbf{A} =
\left(
\begin{array}{rrr}
 2 & -1 &  0 \\
-1 &  2 & -1 \\
 0 & -1 &  2
\end{array}	
\right) \ ,
\hspace{30mm}
\begin{picture}(0,0)(0,0)
	\setlength{\unitlength}{1mm}
	\put(0,0){\circle{2}}
	\put(1,0){\line(1,0){8}}	
	\put(10,0){\circle{2}}
	\put(11,0){\line(1,0){8}}	
	\put(20,0){\circle{2}}
	\put(-3,-5){${\scriptstyle (1,1)}$}
	\put( 7,-5){${\scriptstyle (2,1)}$}
	\put(17,-5){${\scriptstyle (3,1)}$}
\end{picture}
\label{Cartan:SU4}
\end{equation}
Let us take for example the point $\beta_1$=1, $\beta_2$=$\frac{1}{\sqrt{2}}$, 
$\beta_3$=1. The vectors $\vec{\beta}_\nu$ defined in (\ref{N:vertex}) then
reproduce all the simple and some of the other roots of 
$\mathfrak{su}{\scriptstyle (4)}$:
\begin{equation}
	\begin{aligned}
\vec{\beta}_0 & = \vec{\beta}_{000} = 
\left(  \frac{1}{\sqrt{2}},  1,  \frac{1}{\sqrt{2}} \right) =  \alpha_1 \ ,
&~~~& \vec{\beta}_7=\vec{\beta}_{111} 			    = -\alpha_1 \ ; \\
\vec{\beta}_1 & = \vec{\beta}_{001} = 
\left(  \frac{1}{\sqrt{2}},  1, \frac{-1}{\sqrt{2}} \right) 
	 =  \alpha_1{\rm +}\alpha_2{\rm +}\alpha_3 \ ,
&~~~& \vec{\beta}_6=\vec{\beta}_{110}
	 = {\rm -}\alpha_1{\rm -}\alpha_2{\rm -}\alpha_3 \ ; \\
\vec{\beta}_2 & = \vec{\beta}_{010} = 
\left(  \frac{1}{\sqrt{2}}, -1,  \frac{1}{\sqrt{2}} \right) = -\alpha_3 \ ,
&~~~& \vec{\beta}_5=\vec{\beta}_{101}			    =  \alpha_3 \ ; \\
\vec{\beta}_3 & = \vec{\beta}_{011} = 
\left(  \frac{1}{\sqrt{2}}, -1, \frac{-1}{\sqrt{2}} \right) =  \alpha_2 \ ,
&~~~& \vec{\beta}_4=\vec{\beta}_{100}			    = -\alpha_2 \ .
	\end{aligned}
\label{Roots:SU4}
\end{equation}
We see that the exponential form (\ref{N:vertex}) of the 3-cosine perturbation 
contains all $\mathfrak{su}{\scriptstyle (4)}$ generators corresponding to
the simple roots $\alpha_1$, $\alpha_2$, and $\alpha_3$ and their opposite 
negative roots as well as the generators corresponding to the root 
$\alpha_1$+$\alpha_2$+$\alpha_3$ and its opposite. The generators 
corresponding to the remaining roots of $\mathfrak{su}{\scriptstyle (4)}$, 
$\alpha_1$+$\alpha_2$, $\alpha_2$+$\alpha_3$, and their opposite negative 
roots, as well as the 3 Cartan generators, are generated under 
renormalization.

\subsection*{C $~~~$ $\mathfrak{so}{\scriptstyle (8)}$ Current-Current 
Perturbation for $N$=4}

%%%%%%%%%% NUMBERING EQUATIONS WITHIN APPENDIX C %%%%%%%%% 
\makeatletter\@addtoreset{equation}{section}\makeatother 
\setcounter{equation}{0}
\def\theequation{C.\arabic{equation}} 
%%%%%%%%%%%%%%%%%%%%%%%%%%%%%%%%%%%%%%%%%%%%%%%%%%%%%%%%% 

	Here we provide some details of our calculation, showing that the 
4-cosine model can be written as a $\mathfrak{so}{\scriptstyle (8)}$
current-current perturbation at the 16 points on the marginal 4-sphere 
(\ref{Manifold}) specified by (\ref{Points4}).

	The Cartan matrix and the Dynkin diagram of 
$\mathfrak{so}{\scriptstyle (8)}$ are:
\begin{equation}
\hspace{-25mm}
	\mathbf{A} =
\left(
\begin{array}{rrrr}
 2 & -1 &  0  &  0 \\
-1 &  2 & -1  & -1 \\
 0 & -1 &  2  &  0 \\
 0 & -1 &  0  &  2  
\end{array}	
\right) \ ,
\hspace{30mm}
\begin{picture}(0,0)(0,0)
	\setlength{\unitlength}{1mm}
	\put(0,0){\circle{2}}
	\put(1,0){\line(1,0){8}}	
	\put(10,0){\circle{2}}
	\put(11,0){\line(1,0){8}}	
	\put(20,0){\circle{2}}
	\put(10,1){\line(0,1){8}}	
	\put(10,10){\circle{2}}
	\put(-3,-5){${\scriptstyle (1,1)}$}
	\put( 7,-5){${\scriptstyle (2,2)}$}
	\put(17,-5){${\scriptstyle (3,1)}$}
	\put(13, 9){${\scriptstyle (4,1)}$}
\end{picture}
\label{Cartan:SO8}
\end{equation}
At the point $\beta_1=\beta_2=\beta_3=\beta_4=\frac{1}{\sqrt{2}}$, for 
example, the vectors $\vec{\beta}_\nu$ reproduce all the simple and some of 
the other roots of $\mathfrak{so}{\scriptstyle (8)}$:
\begin{equation}
	\begin{aligned}
\vec{\beta}_0 & = \vec{\beta}_{0000} = 
\frac{1}{\sqrt{2}} ( 1, 1, 1, 1)          &=& \ \alpha_2
&~~~& = -\vec{\beta}_{1111} = -\vec{\beta}_{15} \ ; \\
\vec{\beta}_1 & = \vec{\beta}_{0001} = 
\frac{1}{\sqrt{2}} ( 1, 1, 1,-1) 
	&=& \ \alpha_1{\rm +}2\alpha_2{\rm +}\alpha_3{\rm +}\alpha_4
&~~~& = -\vec{\beta}_{1110} = -\vec{\beta}_{14} \ ; \\
\vec{\beta}_2 & = \vec{\beta}_{0010} = 
\frac{1}{\sqrt{2}} ( 1, 1,-1, 1)          &=& \ -\alpha_4
&~~~& = -\vec{\beta}_{1101} = -\vec{\beta}_{13} \ ; \\
\vec{\beta}_3 & = \vec{\beta}_{0011} = 
\frac{1}{\sqrt{2}} ( 1, 1,-1,-1) 
	&=& \ \alpha_1{\rm +}\alpha_2{\rm +}\alpha_3
&~~~& = -\vec{\beta}_{1100} = -\vec{\beta}_{12} \ ; \\
\vec{\beta}_4 & = \vec{\beta}_{0100} = 
\frac{1}{\sqrt{2}} ( 1,-1, 1, 1)          &=& \ -\alpha_3
&~~~& = -\vec{\beta}_{1011} = -\vec{\beta}_{11} \ ; \\
\vec{\beta}_5 & = \vec{\beta}_{0101} = 
\frac{1}{\sqrt{2}} ( 1,-1, 1,-1) 
	&=& \ \alpha_1{\rm +}\alpha_2{\rm +}\alpha_4
&~~~& = -\vec{\beta}_{1010} = -\vec{\beta}_{10} \ ; \\
\vec{\beta}_6 & = \vec{\beta}_{0110} = 
\frac{1}{\sqrt{2}} ( 1,-1,-1, 1) 
	&=& \ {\rm -}\alpha_2{\rm -}\alpha_3{\rm -}\alpha_4
&~~~& = -\vec{\beta}_{1001} = -\vec{\beta}_9 \ ; \\
\vec{\beta}_7 & = \vec{\beta}_{0111} = 
\frac{1}{\sqrt{2}} ( 1,-1,-1,-1)         &=& \ \alpha_1
&~~~& = -\vec{\beta}_{1000} = -\vec{\beta}_8 \ .
	\end{aligned}
\label{Roots:SO8}
\end{equation}
Therefore, the 4-cosine perturbation contains all 
$\mathfrak{so}{\scriptstyle (8)}$ generators corresponding to the simple roots 
$\alpha_1$, $\alpha_2$, $\alpha_3$, $\alpha_4$, and their opposite negative 
roots, as well as the generators corresponding to the roots 
$\alpha_1$+$\alpha_2$+$\alpha_3$, $\alpha_1$+$\alpha_2$+$\alpha_4$, 
$\alpha_2$+$\alpha_3$+$\alpha_4$, 
$\alpha_1$+$2\alpha_2$+$\alpha_3$+$\alpha_4$, and their opposite. The 
remaining generators of $\mathfrak{so}{\scriptstyle (8)}$, corresponding to 
the roots $\alpha_1$+$\alpha_2$, $\alpha_2$+$\alpha_3$, $\alpha_2$+$\alpha_4$, 
$\alpha_1$+$\alpha_2$+$\alpha_3$+$\alpha_4$, and their opposite, as well as
the 4 Cartan generators, are generated under renormalization.

\newpage  
%%%%%%%%%%%%%%%%%%%%%%%%%%%%%%%%%%%%%%%%%%%%%%%%%%%%%%%%%%%%%%%%%%%%%%%%%%%%%%% 
%%%%%%%%%%%%%%%%%%		 BIBLIOGRAPHY		   %%%%%%%%%%%%%%%%%%%% 
%%%%%%%%%%%%%%%%%%%%%%%%%%%%%%%%%%%%%%%%%%%%%%%%%%%%%%%%%%%%%%%%%%%%%%%%%%%%%%% 

\end{document}